\documentclass{article}
\usepackage{spconf,amsmath,graphicx}

\usepackage{multirow}
\usepackage{graphicx}
\usepackage{amssymb, amsmath, bm, mathtools,array}
\usepackage{textcomp}
\usepackage{hyperref}
\usepackage{verbatim,lipsum}
\usepackage{booktabs}
\usepackage{tcolorbox}
\usepackage{lscape}
\usepackage{subcaption}
\usepackage{multirow}
\usepackage{colortbl}
\usepackage{color}
\usepackage[absolute,overlay]{textpos}

\RequirePackage[normalem]{ulem} 
\RequirePackage{color}\definecolor{BLUE}{rgb}{0,0.20,0.75} 
\RequirePackage{color}\definecolor{BROWN}{RGB}{60,128,49} 

\definecolor{mygreen}{RGB}{1, 183, 183}
\definecolor{mygrey}{RGB}{127, 127, 127}

\newcommand{\bs}[1]{\boldsymbol{#1}}
\title{Estimating the confidence of speech spoofing countermeasure}
\name{Xin Wang\thanks{This study is supported by JST CREST Grants (JPMJCR18A6 and JPMJCR20D3), MEXT KAKENHI Grants (21K17775, 21H04906, 18H04112, 16H06302), and Google AI for Japan program.}, Junichi Yamagishi}
\address{National Institute of Informatics, Japan}
\begin{document}
\ninept
\maketitle
\begin{abstract}
Conventional speech spoofing countermeasures (CMs) are designed to make a binary decision on an input trial. However, a CM trained on a closed-set database is theoretically not guaranteed to perform well on unknown spoofing attacks. In some scenarios, an alternative strategy is to let the CM defer a decision when it is not confident. The question is then how to estimate a CM's confidence regarding an input trial. 
We investigated a few confidence estimators that can be easily plugged into a neural-network-based CM. On the ASVspoof2019 logical access database, the results demonstrate that an energy-based estimator and a neural-network-based one achieved acceptable performance in identifying unknown attacks in the test set. On a test set with additional unknown attacks and bona fide trials from other databases, the confidence estimators performed moderately well, and the CMs better discriminated bona fide and spoofed trials that had a high confidence score. Additional results also revealed the difficulty in enhancing a confidence estimator by adding unknown attacks to the training set.
\end{abstract}
\section{Introduction}
\label{sec:intro}
Advanced voice conversion (VC) and text-to-speech (TTS) technologies make it easy to create a high-quality synthetic voice. 
However, synthetic voices can be misused to attack automatic speaker verification (ASV) systems \cite{evans2013spoofing}, now referred to as a  presentation attack (PA) by the ISO/IEC 30107-1 standard \cite{ISO-IEC-30107-1-PAD-Framework-160115}. They can also be abused to fool humans and have lead to an issue known as deepfakes.  These concerns call for reliable PA and deepfake detection methods.

Most PA and deepfake detection methods, or spoofing countermeasures (CM) in general, are based on a binary classification scheme. Given an input speech trial $\bs{x}_{1:T}$ of length $T$, the CM extracts $N$ frames of acoustic features $\bs{a}_{1:N}$ and computes a score $s\in\mathbb{R}$ to indicate how likely the input trial is bona fide -- a real human voice. It then makes a decision by comparing the score with an application-dependent threshold $\theta_s$. 
Most CMs use deep neural networks (DNNs) to detect artifacts in input trials, and many of them have achieved impressive results on benchmark databases \cite{Nautsch2021}.

\begin{figure}[t]
\centering
\includegraphics[trim=0 590 0 80, clip, width=1\columnwidth]{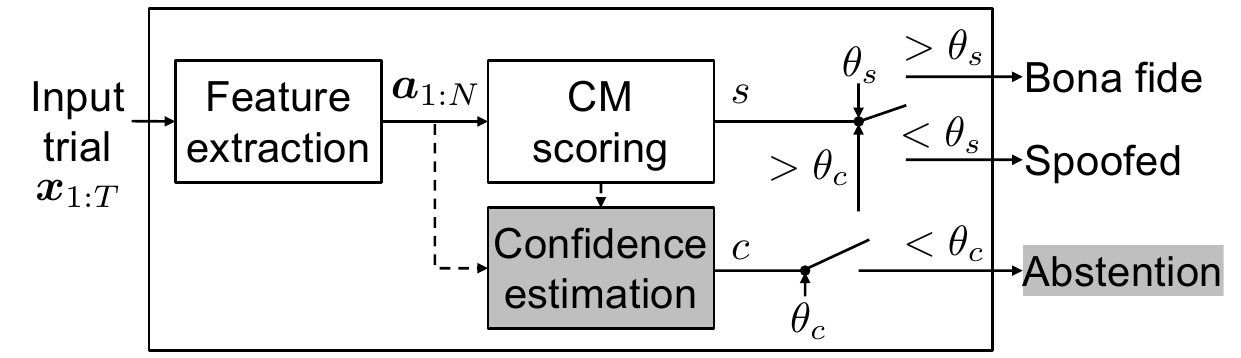}
\vspace{-5mm}
\caption{CM with confidence estimator can opt for abstention. $s$ and $c$ denote CM and confidence scores, respectively. $\theta_s$ and $\theta_c$ denote the threshold for CM classification and abstention, respectively. } 
\vspace{-5mm}
\label{fig:idea}
\end{figure}

However, a CM well trained on a closed-set database is likely to misclassify unseen trials from unknown attacks and unseen bona fide trials from mismatched unknown domains \cite{paul2017generalization,das2020assessing}\footnote{Although benchmark databases such as those from the ASVspoof challenges intentionally keep unknown attacks in the evaluation set, the labels of the evaluation set are released to the public after the challenges. Network architectures and other hyper-parameters of CMs can \emph{unintentionally overfit} the evaluation set.}. These trials are sometimes referred to as ``known-unknown'' and ``unknown-unknown'' in the machine learning field \cite{attenberg2015beat}, and in this study, we simply refer to them as being \emph{unknown} to the CM.
While data augmentation techniques \cite{das21_asvspoof} can make a CM more robust, they cannot cover all unknown conditions.  
Rather than being forced to make a binary decision, a practical CM should abstain from making decisions on trials that are difficult to judge. 
Such a CM is illustrated in Fig.~\ref{fig:idea}. 
The option to abstain is desired when a classification error incurs a high risk, regardless of being a false positive or false negative. If a CM abstains, the input trial can be scrutinized by other CMs or a human expert. 
Although it is not investigated in this study, an active learning strategy can be used to collect the trials annotated by the human expert and fine-tune the CM \cite{settles.tr09}. 

Classification with abstention is an established topic in the machine learning field, and most methods require a separately trained model to decide whether to abstain or not \cite{jiang2018trust}. Other studies treat the task as outlier or out-of-distribution (OOD) data detection. 
They propose augmenting classifiers with a trainable confidence estimator \cite{devries2018learning,Lee2018mdistance} or a non-trainable scoring module \cite{NEURIPS2020_f5496252, hendrycks2016baseline}. OOD data, which is likely to be misclassified by a classifier, usually receives a low confidence score and can be identified.  


Inspired by the aforementioned studies, this study investigates how and whether it is useful to introduce abstention to deep neural network (DNN)-based speech spoofing CMs. On the basis of two high-performance CMs, 
we compared a few confidence estimators on the ASVspoof 2019 logical access (LA) database \cite{WANG2020101114} and an additional evaluation set with \emph{unknown} trials from Voice Conversion Challenges (VCC) \cite{Lorenzo-Trueba2018, Yi2020}. 
The results demonstrate that simply using the probability from a softmax as the confidence score lead to overconfidence, a finding consistent with other studies \cite{guo2017calibration,minderer2021revisiting}.  
An energy-based confidence scoring method \cite{NEURIPS2020_f5496252} and a confidence branch \cite{devries2018learning} achieved an acceptable performance, and both helped the CM to identity \emph{unknown} trials from the VCC test set.
Without making decisions on trials with a confidence score lower than the threshold, the CM achieved a better CM EER on the remaining trials. 
 
The confidence estimators investigated in this study are explained in Sec.~\ref{sec:methods}. The experiments are described in Sec.~\ref{sec:exp}. This paper ends with a conclusion in Sec.~\ref{sec:con}.  Codes and datasets used in this study will be available online\footnote{https://github.com/nii-yamagishilab/project-NN-Pytorch-scripts}.

\begin{textblock*}{19cm}(1cm,1cm) 
\copyright 2021 IEEE. Personal use of this material is permitted. Permission from IEEE must be obtained for all other uses, in any current or future media, including reprinting/republishing this material for advertising or promotional purposes, creating new collective works, for resale or redistribution to servers or lists, or reuse of any copyrighted component of this work in other works.
\end{textblock*}

\section{Confidence estimators}
\label{sec:methods}

\subsection{Max probability from CM}
The first estimator is a simple plug-in to a pre-trained DNN-based CM, for example, a feedforward DNN followed by average pooling, affine transformation, and softmax \cite{wang2021comparative}. 
Suppose the input $\bs{x}_{1:T}$ has been converted into an utterance-level vector $\bs{h}\in\mathbb{R}^{h}$ after average pooling. The output probability given by the affine transformation and softmax output layer can be written as
$P(j | \bs{x}) = \frac{\exp(l_j)}{\sum_{k=1}^{2}\exp(l_k)} = \frac{\exp(\bs{w}_j^\top\bs{h} +\bs{b}_j)}{\sum_{k=1}^{2}\exp(\bs{w}_k^\top\bs{h} +\bs{b}_k)}$,
where $P(j =1 | \bs{x}) $ and $P(j =2 | \bs{x})$ denote the probability of $\bs{x}$ being bona fide and spoofed, respectively, and where $\bs{w}_j$ and $b_j$ are the $j$-th row of the matrix $\bs{W}$ and the bias vector $\bs{b}$ of the affine transformation layer, respectively. 

Given a well-trained CM, the confidence score can be estimated as $c=\max_j P(j | \bs{x})$ \cite{hendrycks2016baseline}. 
Note that, when the CM uses an angular softmax (e.g., \cite{wang2018additive}), the logit is computed as a cosine similarity $l_j = \frac{\bs{w}_j\top\bs{h}}{||\bs{w}_j||\cdot||\bs{h}||}$, and an additional hyper-parameter $\alpha$ is needed to scale the logit and compute $P(j | \bs{x}) = \frac{\exp(\alpha l_j)}{\sum_{k=1}^{2}\exp(\alpha  l_k)}$.

\subsection{Energy-based confidence score}
Given the logits $\bs{l}=[l_1, l_2]^\top$ from a pre-trained CM, 
the second estimator computes an energy-based confidence score as $c = \log\sum_{j=1}^{2}\exp(l_j)$ \cite{NEURIPS2020_f5496252}. 
The value of $c$ is argued to be proportional to the unconditional model likelihood $p_\phi(\bs{x})$, where $\phi$ is the parameter set of the CM. Therefore, \emph{known} attacks are likely to receive a higher score $c$ than \emph{unknown} ones. 

This estimator can be directly used on pre-trained CMs.
When some \emph{unknown} trials are available, it is also possible to re-train the CM with an energy-based training loss \cite{NEURIPS2020_f5496252}, which encourages the confidence scores of \emph{unknown} and \emph{known} trials to be separable.

\subsection{Negative Mahalanobis distance}
Assuming that the utterance-level vectors $\bs{h}$ of trials from the same attack type follow a Gaussian distribution, we can use the negative Mahalanobis distance \cite{Lee2018mdistance} as a confidence estimator. Given the $\bs{h}$ of an input trial, we compute
\begin{equation}
c = - \min_{k} (\bs{h} - \widehat{\bs{\mu}}{_k})^\top \widehat{\bs{\Sigma}}_k^{-1}(\bs{h} - \widehat{\bs{\mu}}_k),
\end{equation}
where $\widehat{\bs{\mu}}_k$ and $\widehat{\bs{\Sigma}}_k$ are the sample mean vector and the covariance matrix of the $k$-th class. 
Note that the class here can be bona fide or any \emph{known} attack in the training set. 
This method has been used for OOD detection \cite{Lee2018mdistance} and CM scoring \cite{chen2015robust}. 
Here, we assume that a trial far away from the \emph{known} classes -- and hence with a smaller score $c$ -- is likely to be \emph{unknown} and misclassified by the CM.  

When some \emph{unknown} trials are available, we can fine-tune the CM to tighten the Gaussian distributions of the \emph{known} classes. This is done in this work using an outlier exposure training loss \cite{hendrycks2019deep}. 

\subsection{Confidence branch}
The fourth confidence estimator adds a trainable module $\mathcal{H}_{\psi}$ to the CM \cite{devries2018learning}. 
It learns to map the vector $\bs{h}$ of the input trial into a confidence score $c=\sigma(\mathcal{H}_{\psi}(\bs{h}))$, where $\sigma(\cdot)$ is the Sigmoid function. The $\mathcal{H}_{\psi}(\cdot)$ is jointly trained with the CM by minimizing the loss over the training data $\{\bs{x}, y\}$:
\begin{equation}
\mathcal{L}_{\psi\phi}(\bs{x}, y) = -\sum_{j=1}^{2} \delta({y = j}) \log \tilde{P}_j - \lambda\log{c},
\label{eq:confbranch}
\end{equation}
where $\tilde{P}_j = c{P}(j | \bs{x}) + (1-c)\delta({y = j})$, $\delta(\cdot)$ is an indicator function, $y\in\{1, 2\}$ is the target label, and $\phi$ denotes the parameter set of the CM scoring module.

When the CM predicts a small confidence score $c$, the value of $\tilde{P}_y$ for the target class is increased, while that of $\tilde{P}_{j\neq{y}}$ is decreased. This means that the CM can ask for more hints from the target label $y$ when it is less confident to classify the input. 
However, the regularization term $- \lambda\log{c}$ prevents the CM from predicting a small $c$ for all trials. Therefore, a well-trained CM is expected to predict a small $c$ only for trials that are difficult to classify.  

We implemented $\mathcal{H}_{\psi}(\bs{h})$ using two linear layers, where the first layer used 128 hidden units and the Tanh activation function.
We followed the official implementation and used a budget mechanism to tune the hyper-parameter $\lambda$. 
We also observed that it is essential to balance the ratio of bona fide and spoofed trials in each mini-batch. 

\subsection{Supervised binary \emph{known-unknown} classifier}
The last estimator in this study is a standalone DNN that predicts the confidence score $\bs{c}$ from input acoustic features. 
It has the same network structure as the CM but the target class is either \emph{known} or \emph{unknown}. 
Bona fide and spoofed trials in the original training set are treated as \emph{known}, and trials from other databases are \emph{unknown}. 

\subsection{Remarks}
The confidence estimators used in this study are summarized in Tab.~\ref{tab:method}. 
We included them because they cover various application scenarios. 
When the CM scoring module has been trained and cannot be fine-tuned, the max-probability, energy-based score, or M-distance can be used. If it is possible to train the CM while only \emph{known} data is available, the confidence branch can be used. If we collect new \emph{unknown} training data, we can choose to build a standalone confidence estimator or update the Gaussian statistics for the M-distance after tuning the CM. 

\begin{table}[t]
\caption{Summary of confidence estimators compared in this study.}
\vspace{-5mm}
\begin{center}
\resizebox{\columnwidth}{!}{
\setlength{\tabcolsep}{2pt}
\begin{tabular}{lllp{3cm}}
\toprule
          & Score range  & \multirow{1}{1.5cm}{Trainable?} & \multirow{1}{3cm}{Use of \emph{unknown} data} \\
\midrule
Max prob.      & $c\in(0.5, 1)$              & No        & Not applicable  \\
Energy score & $c\in\mathbb{R}$    & No        & Usable for CM training \\
Neg. M-dist. & $c \in(-\infty, 0)$        & Trained after CM      & Usable for CM training \\
Conf. branch & $c\in(0, 1)$              & Jointly trained with CM     & Not applicable  \\
Supervised & $c\in(0, 1)$                 & Separately trained    & Required  \\
\bottomrule
\end{tabular}
}
\end{center}
\label{tab:method}
\vspace{-5mm}
\end{table}

\begin{table*}[t]
\caption{Experiment results for CM and confidence scoring. Training and test sets are defined in Tab.~\ref{tab:protocol}. CM scoring components are based on LCNN-LSTM (Section~\ref{sec:recipe}) but have different softmax functions. 
Symbol $\downarrow$ indicates that lower EER, $C_{llr}$, and FPR are better, and $\uparrow$ suggests that higher AUROC and AUPR are better. For visualization, \textbf{cells with better performance have lighter background color}.}
\vspace{-6mm}
\begin{center}
\resizebox{\textwidth}{!}{

\begin{tabular}{cclcccccccccccccccc}
\toprule
\multirow{3}{0.5cm}{\shortstack{Train\\set}}  & \multirow{3}{*}{\shortstack{CM \\ scoring}}   & \multirow{3}{*}{\shortstack{Confidence \\ scoring}} &  &    \multicolumn{6}{c}{Test set E1} &  &  \multicolumn{6}{c}{Test set E2}  \\
\cmidrule{5-10}\cmidrule{12-17}
                                    &  &   &    & $C_{llr}$  & EER & \multicolumn{2}{c}{At TPR=95\%} & AUROC & AUPR &    & $C_{llr}$ & EER & \multicolumn{2}{c}{At TPR=95\%} & AUROC & AUPR  \\
\cline{7-8}\cline{14-15}
                               &  &   & &    $\downarrow$     & $\downarrow$ &  EER $\downarrow$  &  FPR $\downarrow$  & $\uparrow$ &  $\uparrow$ &  &   $\downarrow$     & $\downarrow$ &  EER $\downarrow$  &  FPR $\downarrow$  & $\uparrow$ &  $\uparrow$ \\
\midrule
\multirow{7}{*}{T1} & \multirow{3}{1.cm}{\shortstack{AM \\ softmax}} & Max prob.    & &  \cellcolor[rgb]{0.81, 0.81, 0.81} 0.64 & \cellcolor[rgb]{0.91, 0.95, 0.98} 4.64 &                 -                       &             -                           & \cellcolor[rgb]{0.90, 0.90, 0.90} 0.51 & \cellcolor[rgb]{0.86, 0.86, 0.86} 0.38 & & \cellcolor[rgb]{0.81, 0.81, 0.81} 0.65  & \cellcolor[rgb]{0.87, 0.92, 0.97} 5.50 & -                                      & -                                       & \cellcolor[rgb]{0.91, 0.91, 0.91} 0.51 & \cellcolor[rgb]{0.86, 0.86, 0.86} 0.35\\ 
& & Energy                                                                                & &  \cellcolor[rgb]{0.81, 0.81, 0.81} 0.64 & \cellcolor[rgb]{0.91, 0.95, 0.98} 4.64 & \cellcolor[rgb]{0.93, 0.96, 0.99} 4.25  & \cellcolor[rgb]{0.95, 0.95, 0.95} 80.39 & \cellcolor[rgb]{0.96, 0.96, 0.96} 0.65 & \cellcolor[rgb]{0.97, 0.97, 0.97} 0.63 & & \cellcolor[rgb]{0.81, 0.81, 0.81} 0.65  & \cellcolor[rgb]{0.87, 0.92, 0.97} 5.50 & \cellcolor[rgb]{0.94, 0.97, 0.99} 3.52 & \cellcolor[rgb]{0.94, 0.94, 0.94} 83.34 & \cellcolor[rgb]{0.96, 0.96, 0.96} 0.62 & \cellcolor[rgb]{0.96, 0.96, 0.96} 0.57\\ 
& & Conf. branch                                                                          & &  \cellcolor[rgb]{0.83, 0.83, 0.83} 0.61 & \cellcolor[rgb]{0.95, 0.97, 0.99} 3.72 & \cellcolor[rgb]{0.94, 0.96, 0.99} 3.98  & \cellcolor[rgb]{0.90, 0.90, 0.90} 88.05 & \cellcolor[rgb]{0.99, 0.99, 0.99} 0.75 & \cellcolor[rgb]{1.00, 1.00, 1.00} 0.70 & & \cellcolor[rgb]{0.84, 0.84, 0.84} 0.61  & \cellcolor[rgb]{0.86, 0.91, 0.96} 6.05 & \cellcolor[rgb]{0.87, 0.92, 0.97} 5.50 & \cellcolor[rgb]{0.89, 0.89, 0.89} 89.66 & \cellcolor[rgb]{1.00, 1.00, 1.00} 0.72 & \cellcolor[rgb]{1.00, 1.00, 1.00} 0.68\\ 
\cmidrule{2-17}
& \multirow{4}{1.cm}{plain softmax} &    Max prob.                                  & &  \cellcolor[rgb]{0.95, 0.95, 0.95} 0.45 & \cellcolor[rgb]{0.97, 0.98, 1.00} 3.33 & \cellcolor[rgb]{0.96, 0.98, 1.00} 3.43  & \cellcolor[rgb]{1.00, 1.00, 1.00} 70.98 & \cellcolor[rgb]{1.00, 1.00, 1.00} 0.78 & \cellcolor[rgb]{0.98, 0.98, 0.98} 0.64 & & \cellcolor[rgb]{0.96, 0.96, 0.96} 0.41  & \cellcolor[rgb]{0.87, 0.92, 0.97} 5.55 & \cellcolor[rgb]{0.97, 0.98, 1.00} 2.82 & \cellcolor[rgb]{1.00, 1.00, 1.00} 72.91 & \cellcolor[rgb]{0.99, 0.99, 0.99} 0.70 & \cellcolor[rgb]{0.93, 0.93, 0.93} 0.49\\ 
& &    M-distance                                                                         & &  \cellcolor[rgb]{0.95, 0.95, 0.95} 0.45 & \cellcolor[rgb]{0.97, 0.98, 1.00} 3.33 & \cellcolor[rgb]{0.97, 0.98, 1.00} 3.28  & \cellcolor[rgb]{0.82, 0.82, 0.82} 98.69 & \cellcolor[rgb]{0.91, 0.91, 0.91} 0.55 & \cellcolor[rgb]{0.89, 0.89, 0.89} 0.43 & & \cellcolor[rgb]{0.96, 0.96, 0.96} 0.41  & \cellcolor[rgb]{0.87, 0.92, 0.97} 5.55 & \cellcolor[rgb]{0.92, 0.95, 0.99} 4.14 & \cellcolor[rgb]{0.90, 0.90, 0.90} 88.57 & \cellcolor[rgb]{0.97, 0.97, 0.97} 0.63 & \cellcolor[rgb]{0.95, 0.95, 0.95} 0.52\\ 
& &   Energy                                                                              & &  \cellcolor[rgb]{0.95, 0.95, 0.95} 0.45 & \cellcolor[rgb]{0.97, 0.98, 1.00} 3.33 & \cellcolor[rgb]{0.96, 0.98, 1.00} 3.47  & \cellcolor[rgb]{1.00, 1.00, 1.00} 71.14 & \cellcolor[rgb]{1.00, 1.00, 1.00} 0.79 & \cellcolor[rgb]{1.00, 1.00, 1.00} 0.70 & & \cellcolor[rgb]{0.96, 0.96, 0.96} 0.41  & \cellcolor[rgb]{0.87, 0.92, 0.97} 5.55 & \cellcolor[rgb]{0.97, 0.98, 1.00} 2.85 & \cellcolor[rgb]{1.00, 1.00, 1.00} 72.79 & \cellcolor[rgb]{0.99, 0.99, 0.99} 0.70 & \cellcolor[rgb]{0.93, 0.93, 0.93} 0.49\\ 
& &  Conf. branch                                                                         & &  \cellcolor[rgb]{0.81, 0.81, 0.81} 0.64 & \cellcolor[rgb]{0.95, 0.97, 1.00} 3.60 & \cellcolor[rgb]{0.93, 0.96, 0.99} 4.07  & \cellcolor[rgb]{0.99, 0.99, 0.99} 73.86 & \cellcolor[rgb]{0.99, 0.99, 0.99} 0.76 & \cellcolor[rgb]{1.00, 1.00, 1.00} 0.70 & & \cellcolor[rgb]{0.84, 0.84, 0.84} 0.61  & \cellcolor[rgb]{0.84, 0.90, 0.96} 6.39 & \cellcolor[rgb]{0.89, 0.94, 0.98} 4.95 & \cellcolor[rgb]{0.95, 0.95, 0.95} 81.44 & \cellcolor[rgb]{1.00, 1.00, 1.00} 0.72 & \cellcolor[rgb]{0.98, 0.98, 0.98} 0.61\\ 
\midrule
\multirow{3}{*}{T2}     & \multirow{3}{1.cm}{plain softmax} &  Supervised          & &  \cellcolor[rgb]{0.95, 0.95, 0.95} 0.45 & \cellcolor[rgb]{0.97, 0.98, 1.00} 3.33 & \cellcolor[rgb]{0.97, 0.98, 1.00} 3.22  & \cellcolor[rgb]{0.83, 0.83, 0.83} 97.59 & \cellcolor[rgb]{0.81, 0.81, 0.81} 0.35 & \cellcolor[rgb]{0.81, 0.81, 0.81} 0.29 & & \cellcolor[rgb]{0.96, 0.96, 0.96} 0.41  & \cellcolor[rgb]{0.87, 0.92, 0.97} 5.55 & \cellcolor[rgb]{0.86, 0.91, 0.96} 5.97 & \cellcolor[rgb]{0.81, 0.81, 0.81} 98.17 & \cellcolor[rgb]{0.81, 0.81, 0.81} 0.34 & \cellcolor[rgb]{0.81, 0.81, 0.81} 0.27\\ 
& &  M-distance        &                                                                    &  \cellcolor[rgb]{1.00, 1.00, 1.00} 0.30 & \cellcolor[rgb]{0.92, 0.95, 0.99} 4.34 & \cellcolor[rgb]{0.91, 0.95, 0.98} 4.53 & \cellcolor[rgb]{0.88, 0.88, 0.88} 91.71 & \cellcolor[rgb]{0.90, 0.90, 0.90} 0.52 & \cellcolor[rgb]{0.87, 0.87, 0.87} 0.38  && \cellcolor[rgb]{1.00, 1.00, 1.00} 0.33 & \cellcolor[rgb]{0.87, 0.92, 0.97} 6.52 & \cellcolor[rgb]{0.88, 0.93, 0.97} 6.21 & \cellcolor[rgb]{0.85, 0.85, 0.85} 94.50 & \cellcolor[rgb]{0.92, 0.92, 0.92} 0.53 & \cellcolor[rgb]{0.87, 0.87, 0.87} 0.38\\ 
& &  Energy                                                                          &      &  \cellcolor[rgb]{0.99, 0.99, 0.99} 0.35 & \cellcolor[rgb]{0.89, 0.94, 0.98} 5.06 & \cellcolor[rgb]{0.94, 0.96, 0.99} 3.98  & \cellcolor[rgb]{0.98, 0.98, 0.98} 75.20 & \cellcolor[rgb]{0.98, 0.98, 0.98} 0.73 & \cellcolor[rgb]{0.98, 0.98, 0.98} 0.65 & & \cellcolor[rgb]{0.83, 0.83, 0.83} 0.62  & \cellcolor[rgb]{0.73, 0.84, 0.92} 9.10 & \cellcolor[rgb]{0.84, 0.90, 0.96} 6.47 & \cellcolor[rgb]{0.99, 0.99, 0.99} 74.38 & \cellcolor[rgb]{0.99, 0.99, 0.99} 0.70 & \cellcolor[rgb]{0.93, 0.93, 0.93} 0.49\\ 
\midrule
\multirow{3}{*}{T3}       & \multirow{3}{1.cm}{plain softmax} &  Supervised  &      &  \cellcolor[rgb]{0.95, 0.95, 0.95} 0.45 & \cellcolor[rgb]{0.97, 0.98, 1.00} 3.33 & \cellcolor[rgb]{0.96, 0.98, 1.00} 3.39  & \cellcolor[rgb]{0.82, 0.82, 0.82} 98.44 & \cellcolor[rgb]{0.85, 0.85, 0.85} 0.42 & \cellcolor[rgb]{0.88, 0.88, 0.88} 0.40 & & \cellcolor[rgb]{0.96, 0.96, 0.96} 0.41  & \cellcolor[rgb]{0.87, 0.92, 0.97} 5.55 & \cellcolor[rgb]{0.87, 0.92, 0.97} 5.62 & \cellcolor[rgb]{0.87, 0.87, 0.87} 91.35 & \cellcolor[rgb]{0.93, 0.93, 0.93} 0.55 & \cellcolor[rgb]{0.94, 0.94, 0.94} 0.51\\ 
& &  M-distance                                                                      &      &  \cellcolor[rgb]{0.99, 0.99, 0.99} 0.33 & \cellcolor[rgb]{0.81, 0.88, 0.95} 7.14 & \cellcolor[rgb]{0.80, 0.87, 0.94} 7.25 & \cellcolor[rgb]{0.88, 0.88, 0.88} 91.71 & \cellcolor[rgb]{0.90, 0.90, 0.90} 0.52 & \cellcolor[rgb]{0.87, 0.87, 0.87} 0.38 && \cellcolor[rgb]{0.98, 0.98, 0.98} 0.39 & \cellcolor[rgb]{0.72, 0.83, 0.92} 11.35 & \cellcolor[rgb]{0.72, 0.83, 0.92} 11.26 & \cellcolor[rgb]{0.85, 0.85, 0.85} 94.54 & \cellcolor[rgb]{0.92, 0.92, 0.92} 0.52 & \cellcolor[rgb]{0.87, 0.87, 0.87} 0.38 \\ 
& &  Energy                                                                               & &  \cellcolor[rgb]{1.00, 1.00, 1.00} 0.34 & \cellcolor[rgb]{0.72, 0.83, 0.92} 8.74 & \cellcolor[rgb]{0.76, 0.85, 0.93} 8.12  & \cellcolor[rgb]{0.90, 0.90, 0.90} 88.68 & \cellcolor[rgb]{0.93, 0.93, 0.93} 0.58 & \cellcolor[rgb]{0.91, 0.91, 0.91} 0.47 & & \cellcolor[rgb]{1.00, 1.00, 1.00} 0.34  & \cellcolor[rgb]{0.72, 0.83, 0.92} 9.42 & \cellcolor[rgb]{0.81, 0.88, 0.95} 7.42 & \cellcolor[rgb]{0.94, 0.94, 0.94} 83.71 & \cellcolor[rgb]{0.94, 0.94, 0.94} 0.57 & \cellcolor[rgb]{0.91, 0.91, 0.91} 0.45\\ 
\bottomrule
\end{tabular}
}
\end{center}
\label{tab:result}
\vspace{-5mm}
\end{table*}%

\section{Experiment}
\label{sec:exp}
\subsection{Databases and protocols}
\label{sec:data}
We used three databases: the ASVspoof 2019 LA database \cite{WANG2020101114}, bona fide and TTS trials collected from Blizzard Challenge 2019 (BC19) \cite{wu2019blizzard} and ESPNet \cite{hayashi2020espnet}, and bona fide and VC trials from Voice Conversion Challenge (VCC) 2018 and 2020 \cite{Lorenzo-Trueba2018, Yi2020}. The BC19, ESPNet, and VCC datasets contain more types of spoofed trials than LA, and the sets of speakers are disjoint in all databases. 

To simulate real application scenarios, we prepared different training and test sets, which are listed in Tab.~\ref{tab:protocol}. 
The LA test set was split into two subsets.  \emph{LA test kn.} contains bona fide trials and four spoofing attacks (A08, A09, A16, and A19), and \emph{LA test unk.} contains the rest of the spoofed data in the test set.  
Note that A16 and A19 are \emph{known} because they used exactly the same TTS/VC algorithms as two attackers in the training set. 
A08 and A09 are treated as \emph{known} in this study because they use a TTS framework similar to some spoofing attacks in the training set.

Using T1 and E1 is equivalent to the official protocol of ASVspoof 2019 LA. 
Using E2 simulates a scenario in which some test trials are \emph{unknown}. 
The use of T2 and T3 simulates a scenario in which a small amount of \emph{unknown} trials can be used to train the confidence estimator. 
However, these \emph{unknown} trials are disjoint from those in the test set.

\begin{table}[t]
\caption{List of configurations of training and test sets. Numbers of bona fide and spoofed trials are separated by /.  
}
\vspace{-6mm}
\begin{center}
\resizebox{\columnwidth}{!}{
\begin{tabular}{clll}
\toprule
 & & \emph{Known} trials & \emph{Unknown} trials \\
\midrule
\multirow{3}{0.5cm}{\shortstack{Train \\ set}} & T1 & LA trn. (2,580 / 22,800) & -     \\
& T2 & LA trn.  (2,580 / 22,800) & ESPNet (250 / 2,000)\\ 
& T3 & LA trn. (2,580 / 22,800) & BC19     (100 / 7,525)\\ 
\midrule
\multirow{2}{0.5cm}{\shortstack{Test \\ set}} & E1 & LA test kn.  (7,355 / 19,656) & LA test unk.  (0 / 44,226) \\
& E2 & LA test kn.  (7,355 / 19,656) & VCC   (770 / 49,467) \\
\bottomrule
\end{tabular}
}
\end{center}
\label{tab:protocol}
\vspace{-6mm}
\end{table}%

\begin{figure*}[t]
        \centering
      \begin{subfigure}[t]{0.24\textwidth}
        \includegraphics[width=\textwidth]{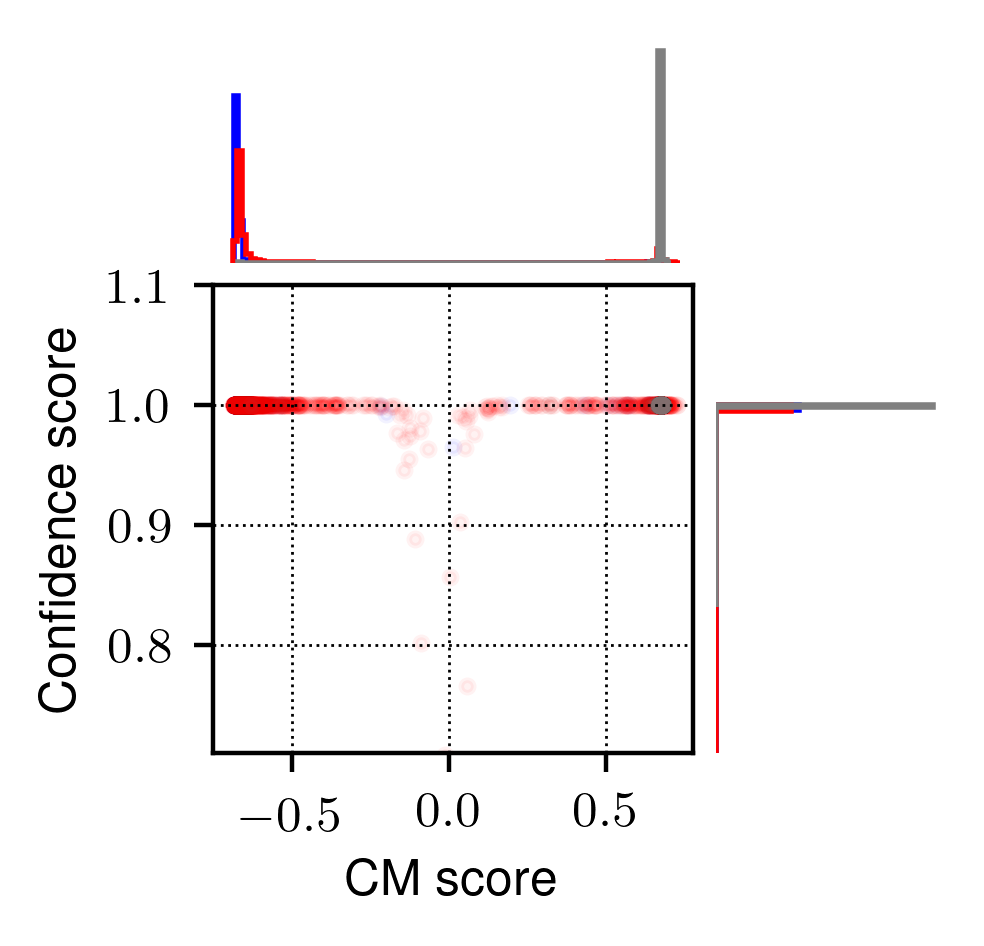}
        \vspace{-6mm}
        \caption{T1-E1, AM, Max prob.}
        \label{fig:sub1}
     \end{subfigure}
      \begin{subfigure}[t]{0.24\textwidth}
        \includegraphics[width=\textwidth]{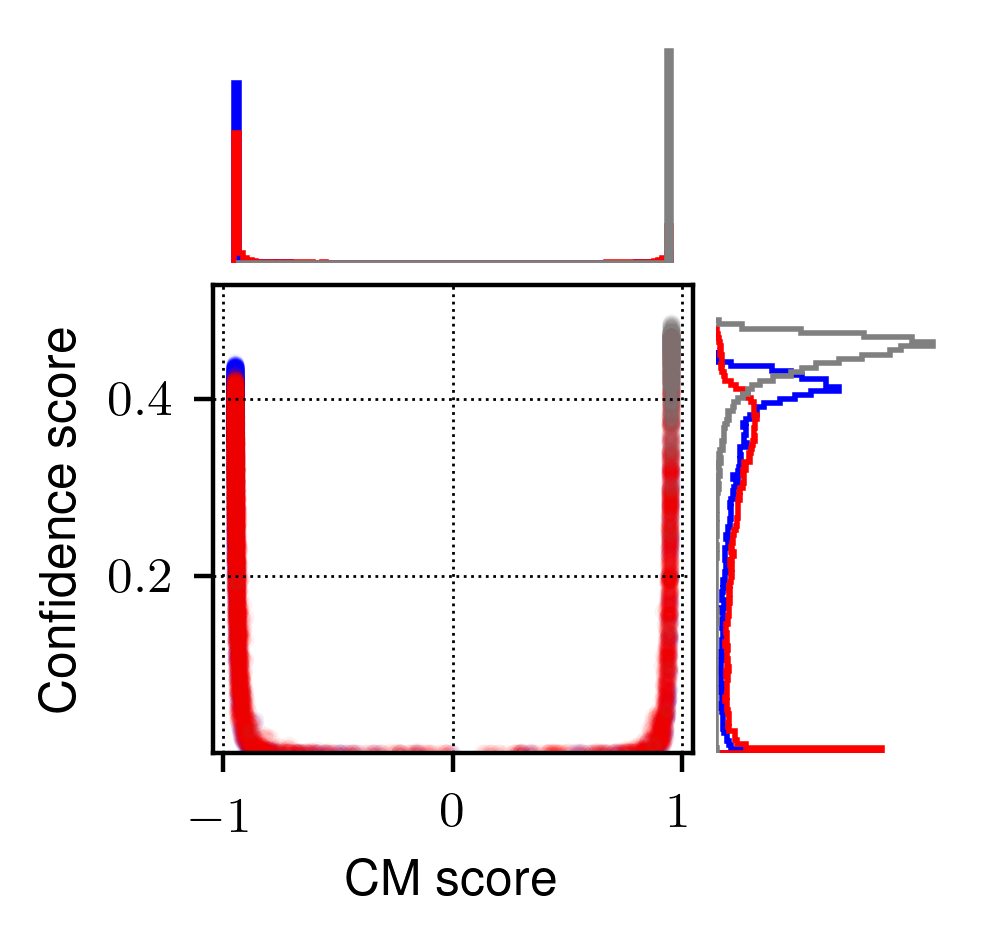}
        \vspace{-6mm}
        \caption{T1-E1, AM, Conf. branch}
        \label{fig:sub2}
     \end{subfigure}
     \begin{subfigure}[t]{0.24\textwidth}
        \includegraphics[width=\textwidth]{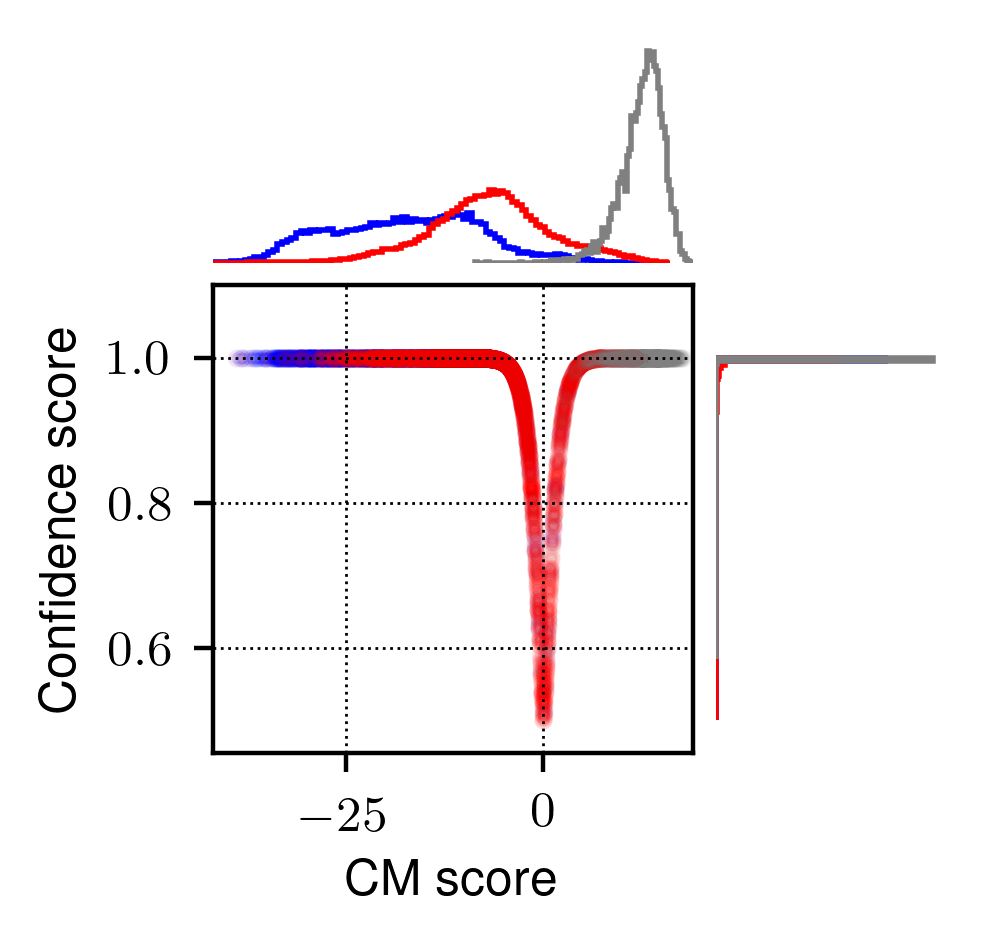}
        \vspace{-6mm}
        \caption{T1-E1, plain, Max prob.}
        \label{fig:sub3}
     \end{subfigure}
      \begin{subfigure}[t]{0.24\textwidth}
        \includegraphics[width=\textwidth]{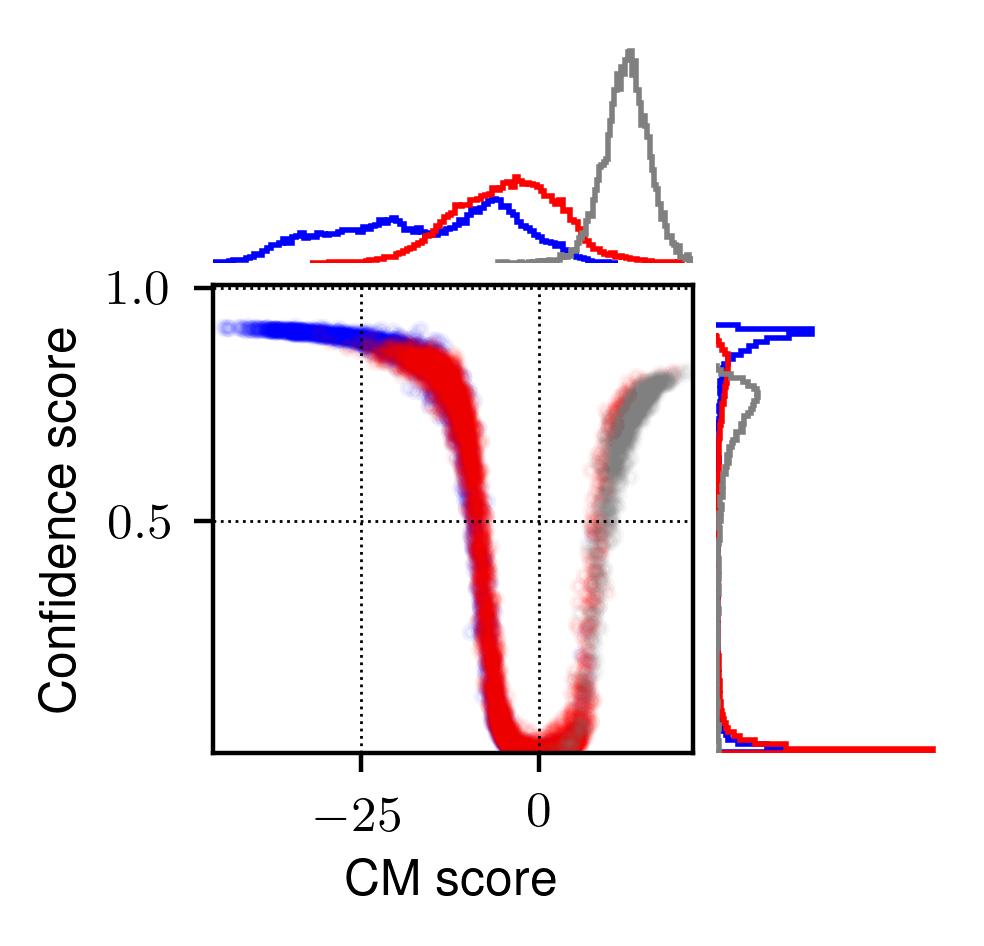}
        \vspace{-6mm}
        \caption{T1-E1, plain, Conf. branch}
        \label{fig:sub4}
     \end{subfigure}
     \hfill
     \begin{subfigure}[t]{0.24\textwidth}
        \includegraphics[width=\textwidth]{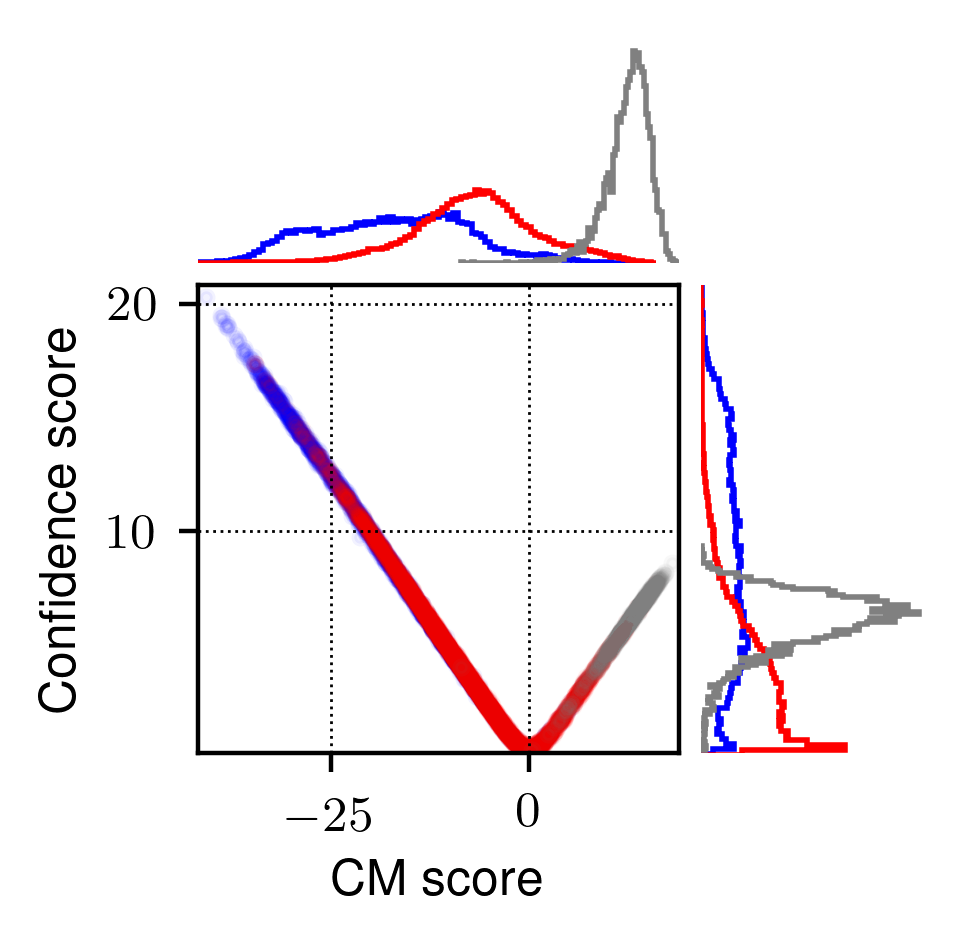}
        \vspace{-6mm}
        \caption{T1-E1, plain, Energy}
        \label{fig:sub5}
     \end{subfigure}
      \begin{subfigure}[t]{0.24\textwidth}
        \includegraphics[width=\textwidth]{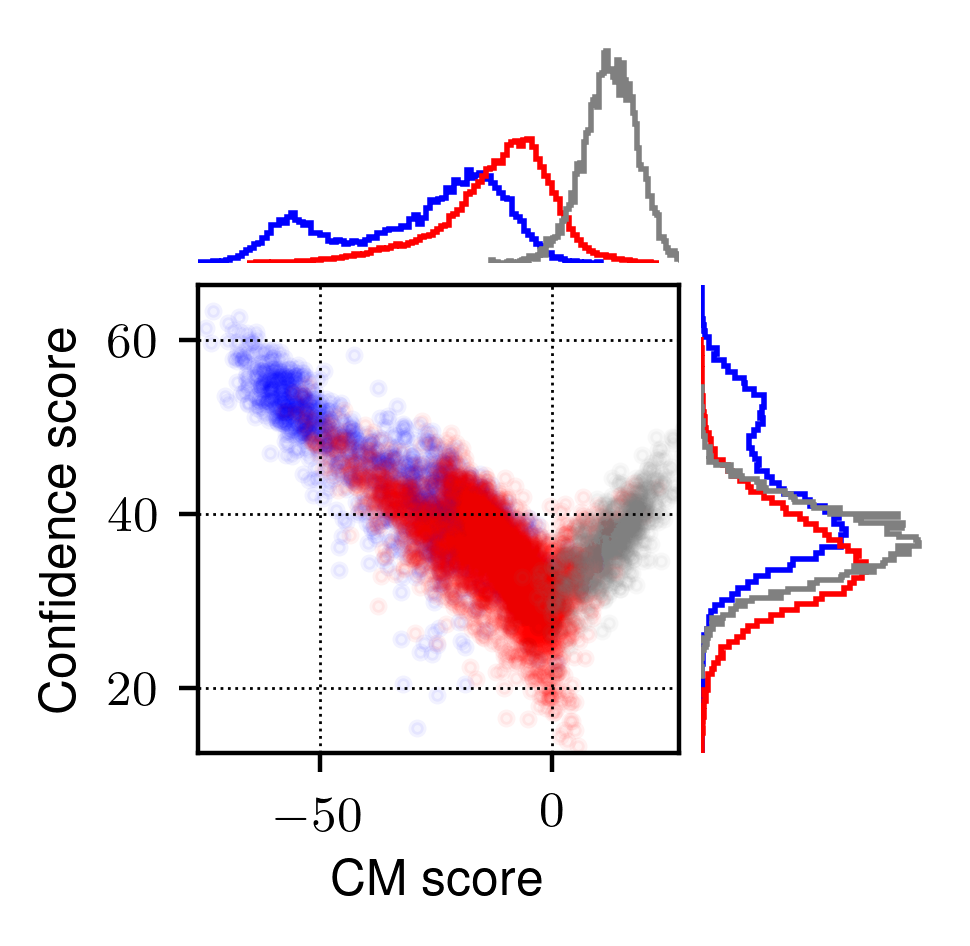}
        \vspace{-6mm}
        \caption{T2-E1, plain, Energy}
        \label{fig:sub6}
     \end{subfigure}
     \begin{subfigure}[t]{0.24\textwidth}
        \includegraphics[width=\textwidth]{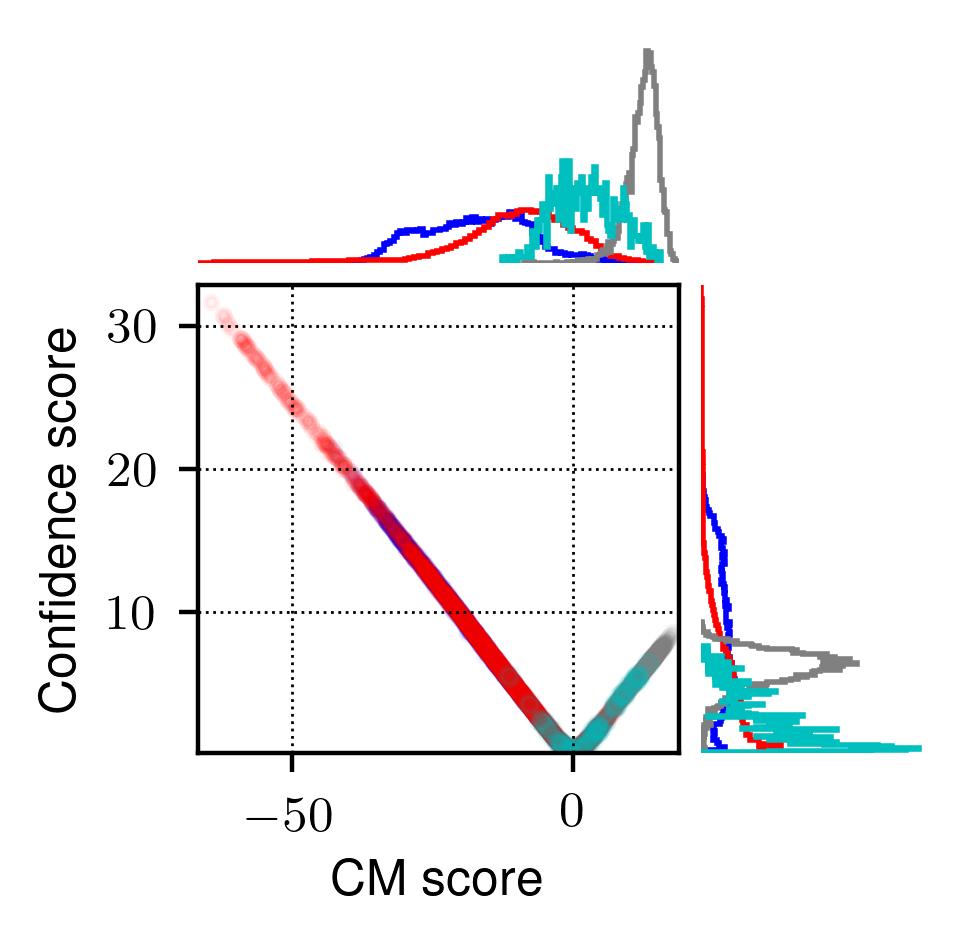}
        \vspace{-6mm}
        \caption{T1-E2, plain, Energy}
        \label{fig:sub7}
     \end{subfigure}
      \begin{subfigure}[t]{0.24\textwidth}
        \includegraphics[width=\textwidth]{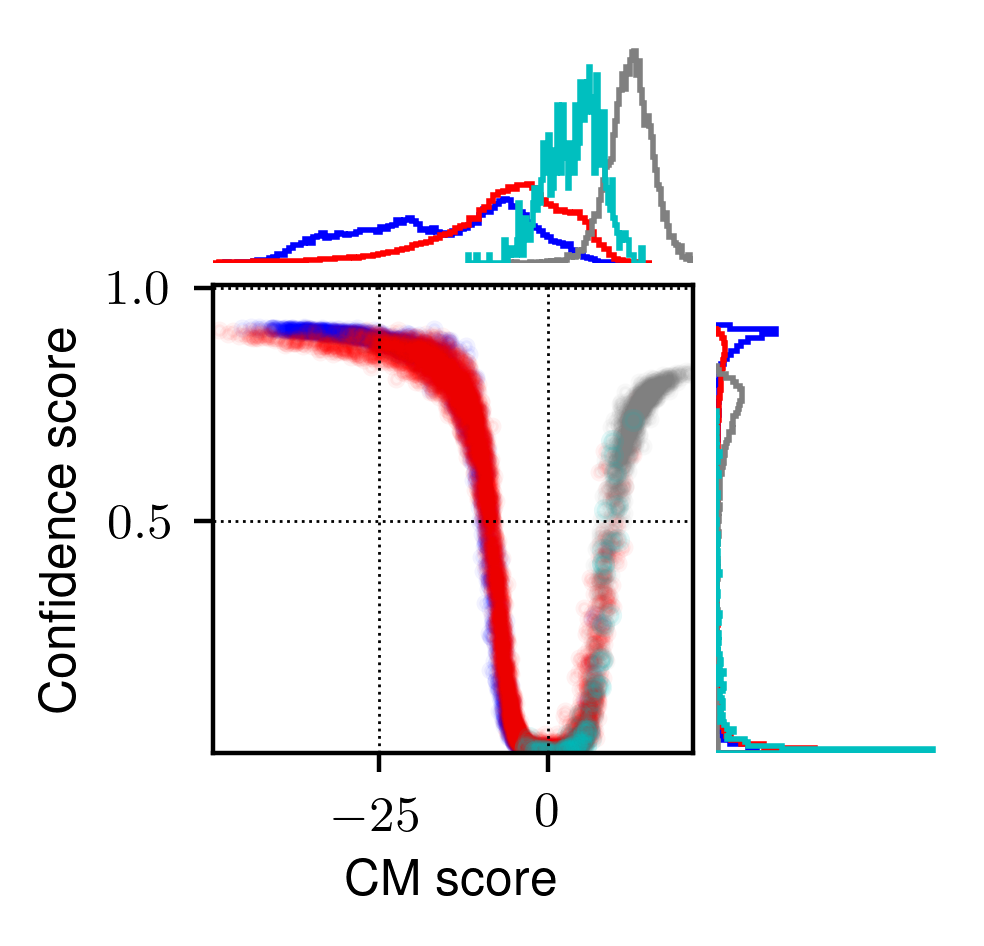}
        \vspace{-6mm}
        \caption{T1-E2, plain, Conf. branch}
        \label{fig:sub8}
     \end{subfigure}
     \vspace{-3mm}
     \caption{Scatter plot and histogram of CM and confidence scores. Each sub-caption lists training-test set, CM with AM or plain softmax, and confidence estimator. \textcolor{mygrey}{\emph{Known} bona fide}, \textcolor{mygreen}{\emph{unknown} bona fide}, \textcolor{blue}{\emph{known} spoofed}, and \textcolor{red}{\emph{unknown} spoofed} trials are in different colors.}
\end{figure*}

\subsection{Model configurations and training recipes}
\label{sec:recipe}
We followed our previous study to configure the CMs since they performed well on the ASVspoof 2019 LA database \cite{wang2021comparative}.
The acoustic features were linear frequency cepstrum coefficients (LFCC) extracted with a frame length of 20ms, a frame shift of 10ms, and a 512-point FFT. The LFCC vector per frame had 60 dimensions, including static, delta, and delta-delta components.
We compared two back-end classifiers in the experiment. While both are based on the light CNN (LCNN) \cite{lavrentyeva2019stc} with two bi-directional LSTM layers and an average pooling layer, one uses an \textbf{plain softmax}, and the other uses an additive margin-softmax (\textbf{AM-softmax}) \cite{wang2018additive} with the hyper-parameter set from \cite{wang2021comparative}. 
These CMs were combined with the confidence estimators for the experiments.

The training recipe was borrowed from our previous study: the Adam optimizer with $\beta_1=0.9, \beta_2=0.999, \epsilon=10^{-8}$ \cite{kingma2014adam},  a mini-batch size of 64, and a learning rate initialized to $3\times10^{-4}$ and halved every ten epochs.  
Each model was trained on an Nvidia Tesla A100 card for three rounds, and the result was averaged. Voice activity detection and feature normalization were not applied. 

\subsection{Evaluation metrics}
\label{sec:metrics}
The following evaluation metrics were used. The first set, including EER and $C_{llr}$ \cite{van2007introduction}, was used to evaluate the CMs in the conventional scenario without discriminating \emph{known} and \emph{unknown} trials. $C_{llr}$ was used because it is a measure of both discrimination and calibration. 

The second set of metrics were for the confidence estimators. 
By treating \emph{known} and \emph{unknown} as positive and negative classes, respectively, we computed the false positive rate (FPR) given the threshold $\theta_c$ for which the true positive (TPR) rate is 95\%. This is used in other studies \cite{hendrycks2016baseline,hendrycks2019deep}. We also computed the area under ROC (AUROC) and the area under the precision-recall curve (AUPR). 

Finally, another EER for the CMs was computed for trials whose confidence score was larger than $\theta_c$ at TPR=95\%. This EER measures how well a CM discriminates bona fide and spoofed trials that the CM is confident about.

\subsection{Results and discussions}
\label{sec:results}
The evaluation results are listed in Tab.~\ref{tab:result}. Note that the CMs with a confidence branch were trained with the loss in Eq.~(\ref{eq:confbranch}), and their $C_{llr}$ and EER were hence different from the others that used a non-trainable confidence estimator.  The CMs with the energy-based confidence estimator were also re-trained when using T2 or T3.

\textbf{Which confidence estimator is effective?} To answer this question, we first focus on the condition using the training set T1, test set E1, and the CMs with the AM softmax. Among the three confidence estimators, the max-probability-based one performed poorly. 
As shown in Fig.~\ref{fig:sub1}, the confidence score was close to the maximum value of 1.0 for most of the \emph{unknown} test trials (in red color), and it was impossible to compute FPR at TPR=95\%.
This indicates that the CM was overconfident, which is consistent with the findings in other studies \cite{guo2017calibration,minderer2021revisiting}.  
In comparison, the energy-based method and the confidence branch produced relatively useful confidence scores. 
Although the FPR at TPR 95\% was higher than 80\% for both methods, the AUROC and AUPR improved. 
Particularly, as shown in Fig.~\ref{fig:sub2}, the confidence branch produced confidence scores that varied for the three types of test trials even though most of the CM scores were either -1 or 1. 

For T1-E1, the max probability, energy-based scoring, and confidence branch improved the AUROC values when the CM used the plain softmax rather than the AM one.  
However, although the results of the max-probability method were close to the other two methods, Fig.~\ref{fig:sub3} shows that many of the \emph{unknown} spoofed trials (in red color) still received a confidence score close to 1.0. 
In contrast, the confidence scores from the other two methods were more dispersed as Figs.~\ref{fig:sub4} and \ref{fig:sub5} show. 
The M-distance-based method was not competitive as the results demonstrate.

The energy-based score and confidence branch were good candidates for the task.
Interestingly, the energy-based confidence scores were highly correlated with the CM scores. This is partially due to the large numeric difference between the logits $\{l_1, l_2\}$. The confidence score becomes $c \approx \max_j(l_j)$ and correlates with the CM score $s = l_1 - l_2$. This also indicates that the logits from the plain softmax contain useful information for confidence estimation.


\textbf{When is the confidence estimation useful for the CM?} 
In the condition T1-E1, the EER at TPR=95\% was not lower than the original EER in most cases. 
The CM cannot better discriminate bona fide and spoofed trials on which the CM is confident.
One possible reason is that the CM has been unintentionally overfitted to the attackers in the LA test set (see the footnote on the 1st page). 
Therefore, to reveal the usefulness of confidence estimation, we need to examine the performance on real \emph{unknown} trials in the test set E2. 
From the results for T1-E2, we observed that the EER at TPR=95\% was lower than the original EER for all confidence estimators except the max-probability-based one for the AM softmax CM. 
As Fig.~\ref{fig:sub7} on the energy-based method shows, the bona fide (in green color) and spoofed (in red color) trials from VCC had smaller confidence scores than those from LA (in grey and blue color). 
This is expected since the trials from VCC were quite different from those in the CM's training set. 
Avoiding making decisions on these low-confidence trials is a reasonable strategy for the CM.

\textbf{Can we improve the confidence estimator if we have some \emph{unknown} spoofing data?} 
A comparison between the metrics across T1, T2, and T3 indicate that it was not effective to use \emph{unknown} training data to fine-tune the CM with an M-distance- or energy-based confidence estimator. In the case of using the energy-based estimator on T2, a comparison between  Fig.~\ref{fig:sub5}  and  \ref{fig:sub6} shows that the confidence score of \emph{unknown} spoofed test trials was pushed towards the \emph{known} ones. 
Although the M-distance's FPR was slightly improved when evaluating on E1, it was higher than 90\%. Note that the M-distance achieved similar FPR, AUROC, and AUPR values for T2-E1 and T3-E1 because the confidence scores were similar.

Last but not least, the estimator trained in a supervised manner produced an AUROC smaller than or around 0.5. One possible reason is that the trials for T2 and T3 were too different from those for T1 and the test set. The confidence estimator learned a simple decision boundary that separated T2 and T3 from T1 but did not generalize to the spoofing attacks in the test set.

\section{Conclusion}
\label{sec:con}
This study investigated speech spoofing CMs that can opt for abstention. This is implemented by augmenting the CMs with a confidence estimator and comparing the confidence score of an input trial against a decision threshold. 
We compared various methods to estimate the confidence score and conducted experiments on a mix of speech databases. The results on the ASVspoof 2019 LA database demonstrated that the energy-based confidence score can be a convenient method for estimating the confidence for pre-trained CMs. The confidence branch is also a potential candidate.
Another experiment with \emph{unknown} spoofed trials from the VCC database showed that the CM can reduce misclassification rate if it can refrain from classifying low-confidence trials.

Future work will investigate other confidence estimation methods, including using calibrated CM score \cite{van2007introduction}. Another direction is to borrow the idea of active learning and add the `unknown' trials with low confidence scores to the CM training set.

\vfill\pagebreak


\bibliographystyle{IEEEbib}
\bibliography{library}

\end{document}